\documentclass[aip,amsmath,amssymb,jap,floatfix,reprint,superscriptaddress,footinbib,showkeys,noeprint]{revtex4-2} 
\usepackage{graphicx}
\usepackage{dcolumn}
\usepackage{bm}
\usepackage[breaklinks,colorlinks,linkcolor={blue},
            citecolor={blue},urlcolor={blue}]{hyperref}
\usepackage[version=4]{mhchem}

\begin{document}

\title{Native point defects in HgCdTe infrared detector material: Identifying {deep centers} from first principles}

\author{Wei Chen}
\affiliation{Institute of Condensed Matter and Nanoscicence (IMCN),
Universit\'{e} catholique de Louvain,
Louvain-la-Neuve 1348, Belgium}
\author{Gian-Marco Rignanese}
\affiliation{Institute of Condensed Matter and Nanoscicence (IMCN),
Universit\'{e} catholique de Louvain,
Louvain-la-Neuve 1348, Belgium}
\author{Jifeng Liu}
\affiliation{Thayer School of Engineering, Dartmouth College,
Hanover, New Hampshire 03755, USA}
\author{Geoffroy Hautier}
\affiliation{Thayer School of Engineering, Dartmouth College,
Hanover, New Hampshire 03755, USA}

\date{\today}

\begin{abstract}
We investigate the native point defects in the long-wavelength infrared (LWIR) detector material \ce{Hg_{0.75}Cd_{0.25}Te} 
using a dielectric-dependent hybrid density functional combined with spin-orbit coupling. Characterizing these point defects 
is essential as they are responsible for intrinsic doping and nonradiative recombination centers in the detector material.
The dielectric-dependent hybrid functional allows for an accurate description of the band gap ($E_g$) for 
\ce{Hg_{1-x}Cd_{x}Te} (MCT) over the entire compositional range, a level of accuracy challenging with standard hybrid functionals.
Our comprehensive examination of the native point defects confirms that cation vacancies
$V_\text{Hg(Cd)}$ are the primary sources of $p$-type conductivity in the LWIR material 
given their low defect formation energies and the presence of a shallow acceptor level ($-$/0) near the valence-band maximum (VBM).
In addition to the shallow acceptor level, 
the cation vacancies exhibit a deep charge transition level (2$-$/$-$) situated 
near the midgap, characteristic of nonradiative recombination centers.
{Our results indicate that Hg interstitial could also be a deep center in the LWIR MCT through a metastable configuration under the Hg-rich growth conditions.
While an isolated Te antisite does not show deep levels, 
the formation of $V_\text{Hg}$--Te$_\text{Hg}$ defect complex introduces a deep acceptor level within the band gap.}
\end{abstract}

\maketitle

\section{Introduction}
Mercury cadmium telluride (MCT) refers to the class of pseudo-binary II--VI semiconductor \ce{Hg_{1-x}Cd_{x}Te},
of which the band gap is direct and tunable across the entire infrared (IR) range via composition engineering~\cite{Norton2002,Rogalski2005}.
{MCTs were initially prepared by bulk crystal growth method in the 1960s but it was proved difficult due to the high Hg vapor pressure at the crystal melting point~\cite{Norton2002}.
Liquid phase epitaxy and, more recently, vapor phase epitaxy techniques such as metal-organic chemical vapor deposition and molecular beam epitaxy are currently the main growth methods for MCTs~\cite{Norton2002}.
Even six decades after the initial proof of concept in IR detection~\cite{Lawson1959},
MCT is still the leading choice in the high-performance detector market
because of its exceptional physical properties, 
among which are high absorption coefficient (1000~cm$^{-1}$)~\cite{Scott1969}, 
high electron mobility ($10^5$~cm$^2$V$^{-1}$s$^{-1}$ at 77~K)~\cite{Yoo1998}, 
and long minority carrier lifetime (up to 1$\mu$s at 77~K)~\cite{Lopes1993,Chen1995,Yoo1998}, compared to competing materials. 
MCT is especially prevalent in long-wavelength IR (LWIR, $>6\mu$m) where its detectivity outcompetes all other technologies~\cite{Vella2021}.

The main recombination mechanisms dominating the dark current, and thereby limiting the ultimate performance of MCT photodiodes, 
include the radiative, Auger, and Shockley-Read-Hall (SRH) nonradiative recombinations.
In particular, the SRH recombination, occurring through deep defect levels within the band gap,
is responsible for minority carrier lifetimes in lightly doped MCT~\cite{Rogalski2005,Kopytko2022}.
At high doping concentrations, these deep centers can facilitate the trap-assisted tunneling of carriers across the junction,
which also contributes to the generation of dark current~\cite{Kopytko2022}.
MCT IR detectors typically require cryogenic cooling systems in order to minimize the diffusion current 
due to thermally generated carriers. High operating temperature (HOT) MCT detectors have recently emerged 
~\cite{Lee2016} which performances are limited by SRH recombination processes.

While there are indications from experiments that native defects are responsible for SRH recombinations in MCT~\cite{Jones1985,Sher1991,Lopes1993}, 
the definitive nature of the SRH centers remains elusive.
The reported positions of SRH centers in MCT span the whole band gap
but their nature is obscure~\cite{Littler1994}.
Numerous experiments suggest the Hg vacancy as the origin of SRH recombinations as Hg-poor grown MCT,
largely $p$-type, 
exhibits considerably shorter minority carrier lifetimes~\cite{Polla1983,Souza1990,Fastow1990,Wijewarnasuriya1994,Nishino2000,Boieriu2006}.
On the other hand, it is well established that the Hg vacancy is a shallow double acceptor~\cite{Vydyanath1981,Gemain2011}.
It is unclear how the dual roles of the Hg vacancy reconcile.
Other proposed candidates for SRH centers include the Hg interstitial, 
which according to resonance ionization spectroscopy measurements gives rise to 
trap levels near midgap in long-wavelength IR (LWIR) $n$-type MCT~\cite{Seiler1991,Littler1992}.
In addition, the Te antisite has been contemplated as a possible origin of the deep levels in undoped $p$-type MCT~\cite{Jones1982,Jones1985}.

The ambiguity in resolving the nature of the deep centers in MCT arises from 
their much lower concentrations compared to the equilibrium carrier concentration of the bulk material.
Many optical spectroscopic techniques are thus ineffective for this purpose 
as optical transitions from deep levels can be obscured, in particular given the narrow band gap of MCTs~\cite{Littler1994}.

The uncertainties associated with the experimental characterizations
emphasize the need for an atomistic understanding of the SRH centers, and more broadly,
the overall defect landscape in MCT from the theoretical viewpoint.
During the early 1990s, a series of pioneering work investigated native point defects in MCT using 
self-consistent first-principles supercell calculations~\cite{Morgan‐Pond1989,Schick1990,Schick1990a,Berding1993,Berding1994,Berding1995},
whereby Hg vacancy were commonly found to be the origin of the $p$-type conductivity of undoped MCT, and Hg interstitial~\cite{Morgan‐Pond1989} and Te antisite~\cite{Berding1995} were among the defects that were considered important
as potential sources of SRH centers.
{A more recent account indicated that Hg vacancy can introduce a deep trap center in addition to the shallow acceptor level~\cite{sun2006}.}
These results, while illuminating, need to be taken with caution because the band gap of MCT was not correctly accounted for
by density functional theory (DFT) within the (semi)-local approximation of the exchange-correlation functional.
Additionally, structural relaxations were either neglected~\cite{Morgan‐Pond1989} or limited to the nearest-neighbor atoms
for only the neutral charge state~\cite{Berding1994}, thereby failing to account for the strong geometrical distortions among 
different charge states that are common to defects in II--VI semiconductors~\cite{Chadi1994,Watkins1996,Lany2001,Lany2004,Kavanagh2021}.
{The lack of a proper treatment of electrostatic finite-size effect for charged defects in a supercell further complicates the interpretation of defect levels and of the defect characteristics.}

In this work, we revisit the native point defects in MCT, specifically the LWIR \ce{Hg_{0.75}Cd_{0.25}Te}, 
by means of a dielectric-dependent hybrid functional~\cite{Chen2018} within the framework of the generalized Kohn-Sham DFT.
As we will show, this dielectric-dependent hybrid functional enables a faithful description of the band gap 
for Hg$_{1-x}$Cd$_{x}$Te over the whole compositional range.
The \ce{Hg_{0.75}Cd_{0.25}Te} is modeled by a special quasirandom structure (SQS)~\cite{Zunger1990}, 
upon which various point defects are incorporated for the determination of the formation energies and charge transition levels.
{Spurious electrostatic interactions for charged defects are accounted for by the correction scheme of Freysoldt \textit{et al}.~\cite{Freysoldt2009,Freysoldt2011}
Our calculations indicate that Hg vacancy, being the dominating native point defects, is both a shallow and deep acceptor.
Hg interstitial can also contribute to the deep levels observed in LWIR MCT through a metastable configuration.
Our results further suggest that the cation vacancy-Te antisite complex is a deep center whereas the Te antisite alone is not.}

The paper is organized as follows.
In Sec.~\ref{sec:method}, we outline the method and computational details.
Section~\ref{sec:pristine} presents the electronic structure for the pristine MCT.
Section~\ref{sec:defect} reports the defect formation energies and charge transition levels for the native point defects and some defect complexes.
The implications of our results are discussed in the context with existing experiments in Sec.~\ref{sec:discussion},
followed by a brief summary and concluding remarks in Sec.~\ref{sec:conclusion}.

\section{\label{sec:method}Method}
\begin{table*}
\caption{\label{tab:pristine}Mixing parameters and calculated band gaps ($E_g$) of Hg$_{1-x}$Cd$_{x}$Te using DD-CAM.
The fraction of Fock exchange in the long range $\alpha_\text{lr}$ corresponds to the inverse of $\epsilon_\infty^\text{expt}$.
The $\epsilon_\infty^\text{expt}$ values are obtained according to Eq.~\eqref{eq:epsilon} except for CdTe~\cite{Lorimor1965}.
For HgTe and CdTe, the macroscopic dielectric constants calculated with DD-CAM are given as $\epsilon_\infty^\text{calc}$.
The negative $E_g$ of HgTe refers to the $\Gamma_6$--$\Gamma_8$ gap due to band inversion [cf.\ Fig.~\ref{fig:gap}(b)].}
\begin{ruledtabular}
\begin{tabular}{lddddd}
             & \multicolumn{1}{c}{HgTe}   
             & \multicolumn{1}{c}{Hg$_{0.75}$Cd$_{0.25}$Te} 
             & \multicolumn{1}{c}{Hg$_{0.5}$Cd$_{0.5}$Te} 
             & \multicolumn{1}{c}{Hg$_{0.25}$Cd$_{0.75}$Te}
             & \multicolumn{1}{c}{CdTe} \\
\hline
$\epsilon_\infty^\text{expt}$ 
             & 15.2   &  11.8                    & 9.5                   & 8.1                       & 7.1 \\
$\mu$ (bohr$^{-1}$) 
             & 0.64   &  0.64                    & 0.65                  & 0.66                      & 0.66 \\
$E_g^\text{calc}$ (eV) 
             & -0.30  &  0.19                    & 0.64                  & 1.14                      & 1.65 \\
$\epsilon_\infty^\text{calc}$ 
             & 14.5   &                          &                       &                           & 7.1 \\
\end{tabular}
\end{ruledtabular}
\end{table*}

\subsection{\label{subsec:ddcam}Dielectric-dependent hybrid functional}
Standard global or range-separated hybrid functionals are based on one or a set of fixed parameters for admixing 
a fraction of Fock exchange with semilocal DFT exchange.
For example, the widely adopted Heyd-Scuseria-Ernzerhof (HSE) hybrid functional admixes 25\% of Fock exchange ($\alpha=0.25$)
in the short range and retains the full DFT exchange in the long range.
The range separation of the Coulomb potential is achieved through an error function 
with an inverse screening length ($\mu$) of 0.106~bohr$^{-1}$~\cite{Heyd2003,Heyd2006}. 
However, as the mixing parameters are material specific, a fixed set of values are inadequate for describing
materials of varying characteristics (e.g., band gap)~\cite{Chen2012}.
The insufficiency of the standard hybrid functional has driven the development of a category of hybrid functionals
that derive the mixing parameters nonempirically by introducing a dielectric dependence (DD)~\cite{Skone2016,Chen2018,Cui2018,Wing2021,Yang2023}.
The DD hybrid functionals have shown remarkable accuracy in predicting the band gaps of a broad range of materials
without the need of any empirical tuning.
Among the various DD schemes available~\cite{Yang2023}, we herein employ the DD-CAM hybrid functional~\cite{Chen2018}, 
a range-separated and dielectric-dependent hybrid functional adopting the Coulomb attenuating method (CAM)~\cite{Yanai2004}.
Specifically, DD-CAM adopts the full Fock exchange in the short range ($\alpha_\text{sr}=1$) and admixes a fraction of Fock exchange
equivalent to the inverse macroscopic dielectric constant in the long range ($\alpha_\text{lr}=\epsilon_\infty^{-1}$).
The screening length can be obtained either by fitting the dielectric function in the long-wavelength limit
or through the Thomas-Fermi (TF) wavevector as proposed by Cui \textit{et al.}~\cite{Cui2018}
\begin{equation}\label{eq:mu}
\mu = \frac{4}{3}\left[ \frac{1}{\gamma} \left( \frac{1}{\epsilon_\infty-1} + 1 \right) \mu_\text{TF}^2 \right]^{\frac{1}{2}},
\end{equation}
where $\gamma=1.563$ is set empirically, 
and the TF wavevector is defined by $\mu_\text{TF}=(3n/\pi)^{1/6}$ where $n$ is the valence electron density.
Generally speaking, the $\mu$ parameter is less critical for the determination of the band gap and it has been shown to 
be well converged near the average value of 0.71 bohr$^{-1}$~\cite{Chen2018}.
As Eq.~\eqref{eq:mu} requires no further calculations apart from $n$ and $\epsilon_\infty$, 
it will be used to determine $\mu$ in our DD-CAM calculations.

\subsection{\label{subsec:comp}Computational details}
\subsubsection{Pristine MCT}
The LWIR MCT is modeled by a 64-atom supercell which contains 8 Cd and 24 Hg atoms, giving rise to the stoichiometry \ce{Hg_{0.75}Cd_{0.25}Te}.
{Since MCTs are known to form in solid solutions~\cite{Longshore2002}, we assume a fully disordered cation sublattice and generate the supercell using the SQS method~\cite{Zunger1990}
as implemented in \texttt{mcsqs} of the \textsc{atat} code~\cite{vandeWalle2013}.}
We take into account the correlations of pair clusters up to the sixth nearest neighbor (NN) and triplet clusters up to the second NN
in the zinc-blende structure.
The resultant cubic SQS supercell reproduces the ideal correlations for all clusters except for the pair cluster at the fourth NN.

All hybrid-functional calculations are performed with the \textsc{vasp} code 
using projector-augmented wave (PAW) potentials~\cite{Kresse1996,Kresse1996a}. 
The $4s$ and $5s$ semicore states are included for Cd and Hg, respectively.
The cutoff energy for the plane-wave basis set is 500~eV.
A $3\times3\times3$ $\Gamma$-centered \textbf{k}-point mesh~\cite{Monkhorst1976} is used for the 64-atom supercell.
For the primitive cell of CdTe and HgTe a $8\times8\times8$ $\Gamma$-centered mesh is used.
The effect of spin-orbit coupling (SOC) is taken into account for single-point total energy and electronic structure, 
as will be applied throughout the study.

A key parameter for DD-CAM calculations is the macroscopic dielectric constant $\epsilon_\infty$.
However due to the size of the SQS supercell, an explicit evaluation of the dielectric constant would be computationally too costly.
Here we take advantage of the quadratic dependence of $x$ for the $\epsilon_\infty$ of Hg$_{1-x}$Cd$_{x}$Te~\cite{Rogalski2005}
\begin{equation}\label{eq:epsilon}
 \epsilon_\infty(x) = 15.2-15.6x+8.2x^2.
\end{equation}
The use of the experimental dielectric constant for Hg$_{1-x}$Cd$_{x}$Te is justified by the fact that the dielectric constants
of CdTe and HgTe, when determined self-consistently with DD-CAM, are in good agreement with experiments (cf.\ Sec.~\ref{sec:pristine}).
This also is in line with the general observation that the dielectric constants predicted on top of DD-CAM 
are highly reliable~\cite{Chen2018}, and as such using the experimental dielectric constant becomes a viable choice.

The lattice constant of Hg$_{1-x}$Cd$_{x}$Te has been shown to deviate from the Vegard's law~\cite{Vegard1921}.
Instead it varies nonlinearly with respect to $x$ as (in \AA)~\cite{Higgins1989}
\begin{equation}\label{eq:a0}
a_0(x) = 6.4614 + (8.4x + 11.68x^2 - 5.7x^3)\times10^{-3},
\end{equation}
which leads to $a_0=6.464$~{\AA} for \ce{Hg_{0.75}Cd_{0.25}Te},
6.46~{\AA} for HgTe, and 6.48~{\AA} for CdTe.
These experimental lattice constants are used in our calculations.

\begin{figure*}[ht]
\includegraphics{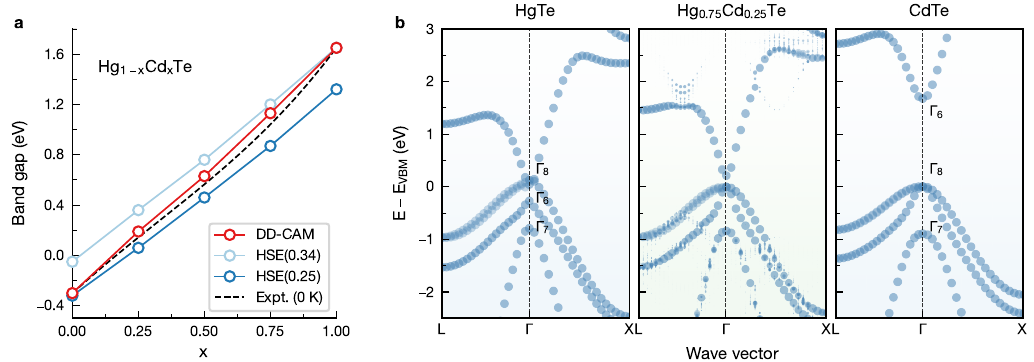}
\caption{\label{fig:gap}(a) Fundamental band gap of \ce{Hg_{1-x}Cd_{x}Te} obtained with HSE and DD-CAM as a function of $x$. The experimental band gap is given by Eq.~\eqref{eq:expt_gap}. The band gap refers to the $\Gamma_6-\Gamma_8$ separation, which becomes negative for HgTe. 
(b) (Effective) band structures of HgTe, \ce{Hg_{0.75}Cd_{0.25}Te}, and CdTe obtained with DD-CAM. 
The symmetries of band-edge states are given in the double-group notation for HgTe and CdTe.
}
\end{figure*}

\subsubsection{Defect computations}
All defect computations are performed with a 216-atom SQS supercell generated by the SQS method using the same clusters as for the 64-atom one.
A fcc-like $\mathbf{k}$-point mesh is used with the 216-atom supercell, resulting in four $k$ points in the Brillouin zone.
Compared to a regular $2\times2\times2$ mesh, the special mesh enables a fully converged total energy within 0.05~meV/atom
at a significantly reduced computational cost.
The band gap is also found to be identical to that of the 64-atom SQS.

The defect formation energy at charge state $q$ is calculated as the total-energy difference given by
\begin{equation}\label{eq:eform}
\Delta E_\text{form}^D(q) = E_\text{tot}^D(q) - E_\text{tot}^\text{bulk} 
- \Sigma_i n_i \mu_i + qE_\text{F} + E_\text{corr},
\end{equation}
where $E_\text{tot}^{D}(q)$ is the total energy of the supercell containing the defect,
$E_\text{tot}^\text{bulk}$ the total energy of the pristine bulk supercell,
$\mu_i$ the chemical potential of the defect species (not to be confused with the screening length), 
$n_i$ the number of atoms added to (or removed from) the supercell,
$E_\text{F}$ the Fermi energy referenced to the VBM ($E_\text{VBM}$),
and $E_\text{corr}$ the electrostatic finite-size corrections of Freysoldt \textit{et al.}~\cite{Freysoldt2009,Freysoldt2011}
using the low-frequency (ion-relaxed) dielectric constant ($\epsilon_0=17$~\cite{Rogalski2005}).
The thermodynamic charge transition level between the two charge states $q$ and $q'$ is expressed as
\begin{equation}
\varepsilon(q/q') = \frac{E_\text{form}^D(q') - E_\text{form}^D(q)}{q-q'} - E_\text{VBM}.
\end{equation}

The atomic positions are allowed to fully relax until the residual forces are less than 0.01~eV/{\AA}.
The lattice constant is kept fixed during structural optimizations.
{For all native defects,  
we consider the charge states up to $q=\pm 2$.}

{We note that a recent study shows that charge transition levels in CdTe converge slowly as $1/L$,
where $L$ is the size of the supercell~\cite{Chatratin2023}.
The slow convergence primarily concerns shallow defects of extended wavefunctions.
Since the defect-defect interaction scales inversely with the dielectric constant,
we expect such a finite-size effect to diminish for defects in \ce{Hg_{0.75}Cd_{0.25}Te}.
This is corroborated by our calculation of a mercury vacancy in a larger 512-atom SQS supercell,
showing that the formation energy changes only by 6 and 8~meV for $q=0$ and $-1$, respectively,
as compared to the results obtained with the 216-atom supercell.}

\section{\label{sec:pristine}Electronic structure of pristine MCT}
We begin with the electronic structure of the pristine \ce{Hg_{0.75}Cd_{0.25}Te}.
To provide an overall assessment of the accuracy of the DD-CAM method for MCT over the entire compositional range,
we additionally include HgTe ($x=0$) and CdTe ($x=1$) as well as two other compositions at $x=0.5$ and $x=0.75$.
For $x=0.5$ a separate SQS supercell is generated using the same clusters as described in Sec.~\ref{subsec:comp}.
Table~\ref{tab:pristine} lists the experimental macroscopic dielectric constants upon which the 
mixing parameters ($\alpha_\text{lr}$ and $\mu$) are determined.
The ensuing DD-CAM band gaps are given in Table~\ref{tab:pristine} and graphically 
in Fig.~\ref{fig:gap}(a) in comparison to the experimental values according to the empirical expression of Hansen \textit{et al.}~\cite{Hansen1982}
\begin{equation}\label{eq:expt_gap}
E_g^\text{expt}(x) = -0.302 + 1.93x - 0.81x^2 + 0.832x^3.
\end{equation}
{Equation~\ref{eq:expt_gap} is a best fit of the data obtained from various optical absorption and magneto-optic experiments~\cite{Hansen1982}.
While the original expression takes an additional term 
accounting for temperature dependence down to 4.2~K, its effect is negligible at low temperatures (less than 3~meV at 5~K).
The temperature dependence is hence neglected in Eq.~\ref{eq:expt_gap} to approximate the band gaps at 0~K.}
The DD-CAM band gaps are in remarkably good agreement with experiments for all the compositions considered hereby.
We also calculate the dielectric constant of CdTe and HgTe using the DD-CAM Hamiltonian in order to assess to what
extent the experimental dielectric constant can be reproduced.
The dielectric response is evaluated by the finite electric field approach~\cite{Souza2002,Umari2002} 
on a $12\times12\times12$ $\mathbf{k}$-point mesh.
As expected, the calculated dielectric constants shown in Table~\ref{tab:pristine} agree favorably with experiments.
This justifies the practice of using experimental dielectric constant in DD-CAM calculations.

Compared to DD-CAM, hybrid functionals with fixed mixing parameters are inadequate in describing the band gap of MCT 
over the entire compositional range.
This is clearly illustrated in Figure~\ref{fig:gap}(a) where the HSE hybrid functional
with a fixed $\alpha$ parameter fails to track the evolution of the band gap with varying $x$. 
Indeed, the optimal $\alpha$ value differs significantly for CdTe and HgTe,
highlighting the need for empirical tuning specific to each stoichiometry for conventional hybrid functionals.

Figure~\ref{fig:gap}(b) shows the band structure of Hg$_{0.75}$Cd$_{0.25}$Te together with those of HgTe and CdTe
along $L$--$\Gamma$--$X$.
Because of the loss of the translational symmetry, the band structure of the Hg$_{0.75}$Cd$_{0.25}$Te supercell is 
unfolded into the Brillouin zone of the zinc-blende primitive cell using spectral decomposition~\cite{Popescu2010}.
The effective band structure retains a highly dispersive and $s$-like conduction-band minimum (CBM),
characteristic of the $\Gamma_6$ symmetry for the CBM of CdTe.
The fourfold degenerate valence-band maximum (VBM) is reminiscent of the VBM of CdTe which is of the $\Gamma_8$ symmetry
and of Te-$5p$ characters.
The spin-orbit splitting between the VBM and the lower-lying light-hole state is 0.8~eV at $\Gamma$, 
which is also close to the $\Gamma_8$--$\Gamma_7$ splitting (0.9~eV) as for CdTe.

Our results indicate that Hg$_{0.75}$Cd$_{0.25}$Te preserves the overall band-structure characteristics 
of CdTe albeit bearing a narrow band gap and a noticeably steeper curvature of the conduction band near $\Gamma$.
The latter is indicative of a small electron effective mass and is in accord with the exceptionally high electron mobility of LWIR MCT.
We estimate an electron mobility of 8000~cm$^2$V$^{-1}$s$^{-1}$ at room temperature 
for Hg$_{0.75}$Cd$_{0.25}$Te from the empirical formula given by Rosbeck \textit{et al.}~\cite{Rosbeck1982}
and Higgins \textit{et al.}~\cite{Higgins1989}
By contrast, the electron mobility of CdTe is 1000~cm$^2$V$^{-1}$s$^{-1}$ at room temperature~\cite{Segall1963,Rode1970}.
We note that the smaller electron effective mass and higher electron mobility associated with heavier cations are typical for II--VI semiconductors~\cite{Rode1970}.

\section{\label{sec:defect}Native defects in MCT}
\begin{figure}
\includegraphics{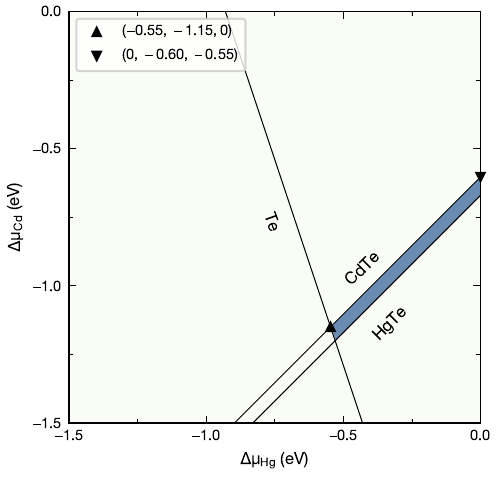}
\caption{\label{fig:pd}Chemical potentials $\Delta \mu_\text{Hg}$ and $\Delta \mu_\text{Cd}$ bound to the formation of \ce{Hg_{0.75}Cd_{0.25}Te} (blue shaded area).
Two growth conditions, namely the Te-rich ($\blacktriangle$) and the Te-poor ($\blacktriangledown$) conditions, are indicated. 
The chemical potential values ($\Delta \mu_\text{Hg}$, $\Delta \mu_\text{Cd}$, $\Delta \mu_\text{Te}$) related to the two growth 
conditions are given (in eV).}
\end{figure}

\begin{figure*}
\includegraphics{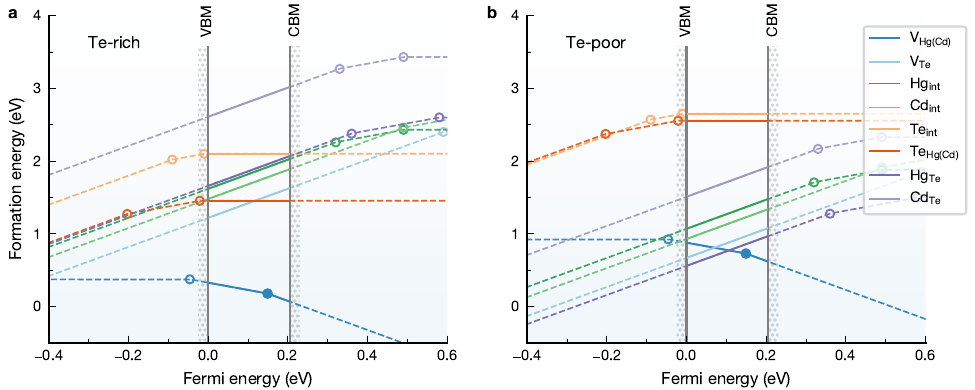}
\caption{\label{fig:defects}Formation energies of native point defects in \ce{Hg_{0.75}Cd_{0.25}Te} 
under (a) Te-rich and (b) Te-poor growth conditions.
The Fermi energy is referenced to the VBM of \ce{Hg_{0.75}Cd_{0.25}Te.}
}
\end{figure*}
 
The native defects considered in our work cover exhaustively all the vacancies ($V_\text{Hg}$, $V_\text{Cd}$, $V_\text{Te}$), 
interstitials (Hg$^\text{int}$, Cd$^\text{int}$, Te$^\text{int}$),
and antisites (Hg$_\text{Te}$, Cd$_\text{Te}$, Te$_\text{Hg}$, Te$_\text{Cd}$).
Cation antisites (Hg$_\text{Cd}$ and Cd$_\text{Hg}$) are not included since they would be electrically inactive.

To arrive at a meaningful defect formation energy, the chemical potentials in Eq.~\eqref{eq:eform} need to be determined
to reflect specific experimental growth conditions. 
To this end, it is convenient to introduce $\Delta \mu_i$ such that 
\begin{equation}
\Delta \mu_i = \mu_i - \mu_i^\text{ref},
\end{equation}
where $\mu_i^\text{ref}$ is the chemical potential reference of the defect species $i$.
Here we take the elementary condensed phases, namely hexagonal Hg and Cd and trigonal Te, for the references.
The chemical potentials are subject to the bounds set by the formation of secondary phases
\begin{align}
\Delta \mu_\text{[Hg,Cd,Te]} \le 0, \\
\Delta \mu_\text{Hg} + \Delta \mu_\text{Te} \le \Delta H_f(\text{HgTe}), \\
\Delta \mu_\text{Cd} + \Delta \mu_\text{Te} \le \Delta H_f(\text{CdTe}), 
\end{align}
where $\Delta H_f$ denotes the formation of enthalpy and is calculated for HgTe and CdTe 
using the aforementioned chemical potential references.
Finally, the stability of Hg$_{0.75}$Cd$_{0.25}$Te requires that 
\begin{equation}
\frac{3}{4}\Delta \mu_\text{Hg} + \frac{1}{4}\Delta \mu_\text{Cd} + \Delta \mu_\text{Te} = \Delta H_f(\text{Hg$_{0.75}$Cd$_{0.25}$Te}).
\end{equation}
For consistency, the enthalpy of formations are calculated with the mixing parameters for Hg$_{0.75}$Cd$_{0.25}$Te (cf.\ Table~\ref{tab:pristine}),
leading to $H_f(\text{HgTe})=-0.27$, $H_f(\text{CdTe})=-0.58$, and $H_f(\text{Hg$_{0.75}$Cd$_{0.25}$Te})=-0.35$, 
all in eV/atom.
At 0~K, the chemical potential region where the stoichiometric Hg$_{0.75}$Cd$_{0.25}$Te is thermodynamically stable 
is narrowly confined by the formation of HgCd and CdTe, as shown by the blue region in the phase diagram in Fig.~\ref{fig:pd}.
At finite temperature the stability of Hg$_{0.75}$Cd$_{0.25}$Te would be enhanced by the configurational entropy due to disorder.
In the present study, we focus on 0~K and thus ignore the vibrational and configurational effects.
Two sets of limiting chemical potentials are selected to represent the Te-rich and Te-poor growth conditions (cf.\ Fig.~\ref{fig:pd})
and are used to determine the defect formation energies.
{Experimentally the growth conditions can be adjusted through the solute content in liquid phase epitaxy 
or the partial pressure in vapor phase expitaxy.
Liquid phase epitaxy growths of MCTs have largely been carried out with Te-rich solutions in favor of a low compensating Hg  pressure and a good control of composition and uniformity~\cite{Schmit1983}.}

Figure~\ref{fig:defects} compiles the formation energy as a function of the Fermi energy for all native defects
under the Te-rich and Te-poor conditions.
We begin with the vacancies and then proceed to the interstitials and antisites.

\begin{figure}
\includegraphics{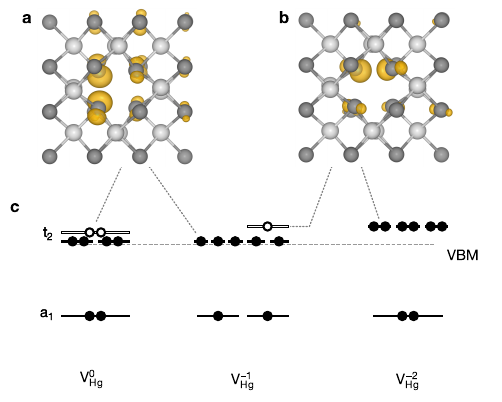}
\caption{\label{fig:vhg}Charge density of $V_\text{Hg}$ associated with the single-particle defect levels in the neutral, 
$-1$, and $-2$ charge states depicted at the $\Gamma$ point. 
Specifically, the charge density labeled (a) is associated with either the unoccupied $t_2$ level of the neutral charge state or the occupied $t_2$ level in the majority channel of the $-1$ charge state, while the charge density labeled (b) is associated with either the unoccupied $t_2$ level in the minority channel of the $-1$ charge state or the fully occupied $t_2$ level in the $-2$ charge state.
The dark gray and light gray balls refer to Te and Hg atoms, respectively.
The charge density is shown by the orange contour with an isovalue of 0.001~$e$/\AA$^3$.
A schematic representation of the single-particle Kohn-Sham (KS) defect levels is given in (c), which assumes the ideal $T_d$ symmetry and depicts the alignment of the $p$-like $t_2$ levels and the $s$-like $a_1$ levels with respect to the VBM. 
The effect of spin-orbit coupling is not taken into account for simplicity.
}
\end{figure}

\begin{figure}
\includegraphics{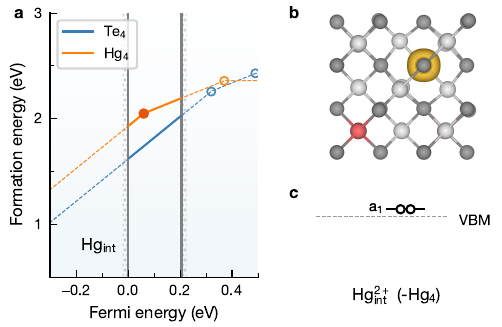}
\caption{\label{fig:ctl_int}{(a) Formation energies of the Hg interstitial in \ce{Hg_{0.75}Cd_{0.25}Te} assuming two tetrahedral coordinations. The Te-rich growth condition is considered.
(b) Charge density at the $\Gamma$ point associated with the lowest unoccupied state of Hg$_\text{int}^{2+}$. The contour isovalue is 0.002~$e$/\AA$^3$.
The dark gray, light gray, and red balls refer to Te, Hg, and Cd atoms, respectively.
A schematic representation of this single-particle defect levels is given in (c).}}
\end{figure}

\subsection{\label{subsec:vacancy}Cation vacancies}
In MCT, the mercury vacancy ($V_\text{Hg}$) is introduced by 
removing one Hg atom that is originally bonded to four Te atoms in a tetrahedral coordination
deviating slightly from the perfect $T_d$ symmetry due to cation disorder.
Upon relaxation of the neutral $V_\text{Hg}^0$, the four neighboring Te atoms move towards the vacancy 
by 0.4~\AA\ with respect to their ideal lattice positions, 
thus maintaining the $T_d$ symmetry~\footnote{More precisely the symmetry is pseudo-$T_d$ because of cation disorder.
But since the distortion is small we always refer to the ideal symmetry in the text for brevity.}.
In the tight-binding picture where the nominal valence electrons are taken into account,
the $V_\text{Hg}^0$ can be described by six electrons contributed equally by 
the four Te atoms, which in the $T_d$ configuration occupy a $s$-like $a_1$ level and partially a threefold degenerate 
$p$-like $t_2$ level (cf.\ Fig.~\ref{fig:vhg}).
The $t_2$ level can be filled by up to two more electrons giving rise to $V_\text{Hg}^-$ and $V_\text{Hg}^{2-}$.
We find that, upon electron addition, the Te atoms show negligible displacements away from the vacancy.
The $T_d$ symmetry is thus preserved for both the neutral and the negative charge states.
Unlike in CdTe~\cite{Kavanagh2021}, we do not find any symmetry-lowering distortion that can lead to a 
more stable vacancy configuration.

Since the Te tetrahedral structure remains unchanged regardless of the lattice site at which
the vacancy is created, the formation energy of $V_\text{Hg}$ is not strongly site dependent.
Here, we consider $V_\text{Hg}$ in a Hg-rich environment (cf.\ Fig.~\ref{fig:vhg}) 
whereby the vacancy is stabilized by 0.15~eV more than in a Cd-rich environment.
As shown in Fig.~\ref{fig:defects}, $V_\text{Hg}$ in \ce{Hg_{0.75}Cd_{0.25}Te} features
a shallow $\varepsilon$($-$/0) acceptor level in proximity to the VBM 
as well as a deep $\varepsilon$(2$-$/$-$) acceptor level positioned 0.15~eV above the VBM.
The shallow acceptor level arises from the near-degeneracy of the $t_2$ levels 
in the neutral and $-1$ charge states due to the tetrahedral coordination, 
leading to a marginal separation between the unoccupied and the occupied $t_2$ levels, the
latter being in resonance with the VBM.
The deep level exhibits a more localized character associated with Te-$p$ orbitals 
(cf.\ Fig.~\ref{fig:vhg}).

The cadmium vacancy ($V_\text{Cd}$) is indistinguishable from $V_\text{Hg}$
due to the isostructural and isoelectronic nature of the two cation defects.
Accordingly, the two cation vacancies function as a shallow acceptor while still maintaining a mid-gap acceptor level.

\subsection{Tellurium vacancy}
Removal of one Te atom from the pristine MCT effectively leaves behind two electrons 
that can subsequently be depopulated to create positively charged Te vacancies. 
Similar to the cation vacancies, $V_\text{Te}$ in MCT demonstrates the tetrahedral symmetry with no 
indication of polaronic distortion.
The two electrons, which are supposedly of the $s$-like $a_1$ character, 
hybridize with the conduction band and are delocalized over the whole supercell.
As a result, the charge transition levels are well above the CBM, 
and $V_\text{Te}$ is doubly ionized regardless of the Fermi level.

We note that, in contrast to the cation vacancies, $V_\text{Te}$ adopts a range of
nearest neighbors due to the randomly distributed Hg and Cd atoms.
With the current SQS, the majority of the 32 $V_\text{Te}$-centered tetrahedra 
are characterized by the Hg-rich motifs, with \ce{Hg3Cd} and \ce{Hg4} accounting for 43\% and 31\% 
of them, respectively.
The remaining motifs are comprised of \ce{Hg2Cd2} (20\%) and \ce{HgCd3} (6\%).
The $V_\text{Te}$ is more likely to form within the Hg-rich regions,
and this can be attributed to the substantially lower enthalpy of formation of HgTe
in comparison to CdTe.
In Fig.~\ref{fig:defects} $V_\text{Te}$ specifically refers to the vacancy formed within the \ce{Hg4}
structure, of which the formation energy can be 0.6~eV lower than those formed within Cd-rich motifs.
Nevertheless, no charge transition level of $V_\text{Te}$ is found within the band gap 
irrespective of the formation conditions.

\subsection{Cation interstitials}
Tetrahedral interstitials, typically the most stable configuration for cation interstitials 
in II--VI zinc-blende semiconductors~\cite{Wei2002,Ma2014},
are also preferred in MCT according to our calculations.
Specifically, the cation interstitials can form tetrahedral bonds with either four Te atoms (Te$_4$) or four cationic atoms.
Figure~\ref{fig:ctl_int}(a) shows the formation energy of Hg$_\text{int}$ in both tetrahedral coordinations.
For the cationic coordination, we consider four Hg atoms (Hg$_4$). 
While the neutral Hg interstitial (Hg$_\text{int}$) is slightly more stable within the Hg$_4$ shell,
the positively charged interstitials are more stabilized under the Te$_4$ coordination by up to 0.25~eV, primarily 
through the pronounced Coulomb attractions.
This leads to distinct charge transition levels under the two coordinations as is manifested in Fig.~\ref{fig:ctl_int}(a).
Notably, Hg$_\text{int}$--Hg$_4$ has the donor level ($+$/2$+$) that is located 60~meV above the VBM, 
whereas this donor level becomes extremely shallow for Hg$_\text{int}$--Te$_4$.
The (0/$+$) level is nonetheless always above the CBM.

The ($+$/2$+$) donor level is associated with the $s$-like single-particle level localized at the 
interstitial as illustrated in Fig.~\ref{fig:ctl_int}(b).
The $s$-like state originates from the $a_1$ representation of the $T_d$ symmetry which is consistently preserved
for the cation interstitials in all relevant charge states.
This level is clearly discernible within the band gap for Hg$_\text{int}$--Hg$_4$ [cf.\ Fig.~\ref{fig:ctl_int}(c)],
but it becomes resonant with the conduction band for Hg$_\text{int}$--Te$_4$, 
in accord with the disparities observed in the charge transition levels.

Meanwhile, the donor levels of the Cd interstitial (Cd$_\text{int}$) 
are consistently higher compared to Hg$_\text{int}$.
Consequently, Cd$_\text{int}$ does not possess any defect levels within the band gap under either tetrahedral coordination and remains doubly ionized.

\subsection{Tellurium interstitial}
In contrast to the cation interstitials, tellurium interstitial (Te$_\text{int}$)
is found to be unstable in the high-symmetry tetrahedral coordination.
Instead, it forms a split Te--Te dumbbell that is oriented approximately along the $\langle 100 \rangle$ direction
in the neutral and $+1$ charge state.
The Te--Te bond length is about 2.7~\AA\ and is only marginally contracted in the $+1$ charge state.
When in the $+2$ charge state, Te$_\text{int}$ becomes a hexagonal interstitial, 
with an average Te--Te (Te--Hg) bond length of 2.85 (3.15)~\AA.
Figure~\ref{fig:defects} reveals no deep level associated with Te$_\text{int}$ as the two donor levels are situated below the VBM.

In CdTe, a metastable Te$_\text{int}$ configuration is found where the interstitial sits
in the center of a Te-Te bond along the $\langle 100 \rangle$ direction~\cite{Kavanagh2022}.
The metastable configuration is in effect a saddle point connecting two dumbbell configurations.
We find that such a configuration would always relax to the dumbbell configuration and hence is not likely to play a role in \ce{Hg_{0.75}Cd_{0.25}Te}.

\subsection{Cation antisites}
Cation antisites on the Te lattice sites (Hg$_\text{Te}$ and Cd$_\text{Te}$) 
can be described as a system of four electrons (two connected to the Te vacancy and 
the other two to the Cd atom) in the $a_1^2t_2^2$ electron configuration
according to the $T_d$ symmetry.
A Jahn-Teller distortion is hence expected (except for the +2 charge state)
due to the threefold degenerate $t_2$ level being partially filled.
In effect the neutral cation antisites display a $C_{2v}$ symmetry,
whereas the symmetry is further lowered in the $+1$ charge state.

We calculate the formation energy of the cation antisites in various charge
states and find no charge transition levels within the band gap 
as the they are all above the CBM (cf.\ Fig.~\ref{fig:defects}).
We recall that the donor levels of the cation antisite are 
determined by the $p$-like state stemming from the antisite
and the surrounding cations, for which the single-particle level
remains in resonant with the conduction band irrespective of 
the charge state.
The cation antisites are thereby doubly ionized in \ce{Hg_{0.75}Cd_{0.25}Te}.

\subsection{Tellurium antisite}
Tellurium antisites (Te$_\text{Hg}$ and Te$_\text{Cd}$) are associated with 12 electrons 
based on the analysis of nominal valence electrons.
Under the original $T_d$ symmetry, eight electrons occupy fully the bonding $a_1$ and $t_2$ levels whereas the remaining four 
electrons occupy the antibonding $a_1$ level and partially the $t_2$ level.
The partially occupied $t_2$ level would suggest that the system undergoes a Jahn-Teller distortion away from the ideal $T_d$ symmetry,
similar to Te$_\text{int}$ and the cation antisites.
Indeed, Te antisites in MCT exhibit the $C_{3v}$ symmetry in the neutral and $+1$ charge states, with the Te atom displaced along 
the $\langle 111 \rangle$ direction.
Te antisites in the $+2$ charge state still exhibits the $T_d$ symmetry as the antibonding $t_2$ level is unoccupied.

Figure~\ref{fig:defects} shows that the donor levels of Te$_\text{Hg}$ and Te$_\text{Cd}$ are found below the VBM,
rendering the Te antisites effectively neutral in \ce{Hg_{0.75}Cd_{0.25}Te}.
The absence of donor levels within the band gap is the result of the close proximity of the $p$-like
level of the Te interstitial to the VBM, along with the substantial reduction in formation energy
for the neutral and $+1$ charge states due to the Jahn-Teller distortion.

\subsection{Defect complexes}
{Point defects can also appear in the form of defect complexes whereby the formation energy can be different from the sum of the isolated forms due to charge transfer and Coulomb interactions.
Because of the predominance of the cation vacancy in \ce{Hg_{0.75}Cd_{0.25}Te}, we investigate three $V_\text{Hg}$-related complexes, 
namely the Frenkel defect $V_\text{Hg}$--Hg$_\text{int}$, 
the Schottky divacancy $V_\text{Hg}$--$V_\text{Te}$,
and the vacancy-antisite pair $V_\text{Hg}$--Te$_\text{Hg}$.
All three defect complexes are found to be stable against dissociation into individual defects as indicated by the calculated binding energies shown in Fig.~\ref{fig:complex}(a).}

{Figure~\ref{fig:complex}(b) shows that the Frenkel defect and the divacancy do not introduce any charge transition level within the band gap,
and are therefore electrically inactive in the MCT.
By contrast, the $V_\text{Hg}$--Te$_\text{Hg}$ complex displays a deep acceptor level ($-$/0) at 0.14~eV above the VBM.
This deep level is associated with pronounced $p$-like orbitals localized at the Te antisite as well as the Te atoms surrounding the cation vacancy [cf.\ Fig.~\ref{fig:complex}(b) inset].}

\begin{figure}
\includegraphics{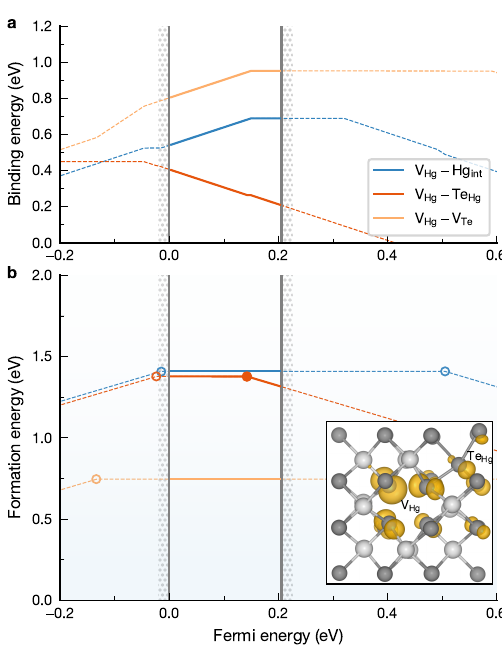}
\caption{\label{fig:complex}{(a) Binding energy of $V_\text{Hg}$-related defect complexes as a function of Fermi level. (b) Formation energy of the defect complexes as a function of Fermi level. The formation energy of $V_\text{Hg}$--Te$_\text{Hg}$ is obtained under the Te-rich conditions, and increases by 1.65~eV under the Te-poor conditions.
The inset illustrates the charge density (at the $\Gamma$ point) associated with the lowest unoccupied state of the neutral $V_\text{Hg}$--Te$_\text{Hg}$ complex.}}
\end{figure}

\section{\label{sec:discussion}Discussion}
Our survey over all possible native point defects indicates that the cation vacancies ($V_\text{Hg}$ and $V_\text{Cd}$) 
are the only acceptors in \ce{Hg_{0.75}Cd_{0.25}Te}.
The cation vacancies are not compensated by any of the donors under Te-rich growth conditions
thanks to the exceptionally low defect formation energy of the vacancies,
which naturally leads to the $p$-type doping as typically observed for as-grown MCT.
Compensation of the vacancy acceptors only occurs under highly Te-poor conditions by the Hg$_\text{Te}$ antisite,
pinning the Fermi level near the midgap (cf.\ Fig.~\ref{fig:defects}).
In either case, MCT is not expected to exhibit $n$-type doping through the native point defects.
This aligns with experimental observations that MCTs are intrinsically $p$-type.~\cite{Lei2015}.

Cation vacancies are generally shallow acceptors in II--VI semiconductors~\cite{Wei2002}.
The fact that cation vacancies act as both shallow acceptors and deep centers in \ce{Hg_{0.75}Cd_{0.25}Te} is intriguing.
The ultrashallow ($-$/0) level is attributed to the near degenerate $t_2$ level of $T_d$ symmetry, 
which in the neutral charge state brings the unoccupied defect state close to the VBM.
As a result, the single-particle defect level is admixed with the bulk resonance to some extent (cf.\ Fig.~\ref{fig:vhg}).
While the degeneracy can be lifted through a Jahn-Teller distortion to give rise to a deeper ($-$/0) acceptor level
as found for the neutral $V_\text{Cd}$ in CdTe~\cite{Kavanagh2021},
we find no such distortion in \ce{Hg_{0.75}Cd_{0.25}Te}, suggesting that the polaron cannot be 
stabilized by the strong screening in the event of a very narrow band gap.
The $(2-/-)$ level, on the other hand, corresponds to the ionization of the completely filled $t_2$ level,
which is localized and sits roughly halfway between the band edges.

The other point defects, all of the donor type, generally do not exhibit any defect level within the band gap of \ce{Hg_{0.75}Cd_{0.25}Te}.
This is at variant with CdTe which is more likely to host deep defect levels of donors such as $V_\text{Te}$~\cite{MenendezProupin2019} due to the much larger band gap.
It is however worth noting that, in the metastable configuration bonded tetrahedrally to four cation atoms, Hg interstitial bears a localized ($+/2+$) donor level located 60~meV above the VBM.
Such a localized defect level could be relevant under Hg-rich growth conditions, making the Hg interstitial more liable to nonradiative carrier capture.
{Using the nudged elastic band method~\cite{Mills1995}, we obtain an energy barrier of 0.25~eV for the neutral Hg interstitial in the metastable configuration to
overcome in order to relax to the ground-state configuration.
It is thus plausible that Hg interstitials can be stabilized in the
metastable configuration, particularly at low temperature and under Hg-rich conditions.}

{Our results show that Te antisite as an isolated point defect is electrically inactive in MCT. 
Nevertheless, the antisite in the form of defect complex when coupled to a neighboring cation vacancy is found to be a possible origin of deep levels observed in MCT.
The acceptor level ($-$/0) nearly overlaps with the (2$-$/$-$) level of $V_\text{Hg(Cd)}$, potentially making it challenging to distinguish between the two defects. 
In Figure~\ref{fig:deeplevels} the charge transition levels of the three deep centers identified from our calculations are summarized.
Notably, the charge transition levels are situated at approximately 0.25$E_g$ and 0.6$E_g$ above the VBM.}

\begin{table}
\caption{\label{tab:expt}Compilation of defect levels observed experimentally in undoped \ce{Hg_{1-x}Cd_{x}Te}.
Defect levels are referenced to the VBM and are expressed in terms of the band gap $E_g$, 
which is given and estimated by Eq.~\eqref{eq:expt_gap}.} 
\begin{ruledtabular}
\begin{tabular}{llllll}
{$x$} & $E_g$ (eV) & Levels ($E_g$) & Assigned defect& Ref. \\
\hline
Donor-like\\
0.215 & 0.08 & 0.42, 0.55         &   -               & \onlinecite{Polla1981} \\
0.24  & 0.13 & 0.5                &   Hg$_\text{int}$ & \onlinecite{Littler1992} \\
0.271 & 0.18 & 0.39, 0.45         &   -               & \onlinecite{Polla1981} \\
0.305 & 0.23 & 0.69, 0.89         &   -               & \onlinecite{Polla1981} \\
0.32  & 0.26 & 0.4                &   -               & \onlinecite{Polla1984} \\
0.2--0.4 & 0.06--0.39 & 0.4, 0.75 &   Te$_\text{Hg}$ & \onlinecite{Jones1982} \\
Acceptor-like\\
0.20  & 0.06 & 0.12, 0.7          &   $V_\text{Hg}$   & \onlinecite{Hoeschl1988} \\
0.216 & 0.09 & 0.15, 0.4          &   $V_\text{Hg}$     & \onlinecite{Gold1986}  \\
0.23  & 0.11 & 0.21               &   $V_\text{Hg}$   & \onlinecite{Sarusi1992} \\
0.224 & 0.10 & 0.07, 0.50         &   $V_\text{Hg}$   & \onlinecite{Bartoli1986} \\ 
\end{tabular}
\end{ruledtabular}
\end{table}

Table~\ref{tab:expt} compiles the experimentally identified defect levels in LWIR MCTs.
Certain experimental characterizations have revealed the coexistence of a very shallow acceptor level and a deeper acceptor level, 
the concentrations of which are highly correlated with the Hg partial pressure~\cite{Vydyanath1981,Bartoli1986,Gold1986,Hoeschl1988}.
The origin of these acceptor levels was believed to be the Hg vacancy, 
and this hypothesis is fully corroborated by our first-principles calculations. 
In particular, the experimentally observed positions linked to the deep V$_\text{Hg}$ acceptor level (ranging from 0.4 to 0.7$E_g$) agree well with our computed level (0.7$E_g$ above the VBM).
Our theoretical results clarifies that the Hg vacancy can be at the same time a shallow acceptor ($-$/0) contributing to $p$-type doping and a deep defect (2$-$/$-$). Our results also indicate that the Cd vacancy behaves similarly than mercury in the mercury-rich alloy. This suggests that control of the Cd chemical potential could be as important as the Hg chemical potential.

Other characterizations have observed donor-like defect levels in $p$-type MCTs~\cite{Polla1981,Jones1982,Polla1984}.
In particular, Jones \textit{et al.}\ reviewed the deep-level transient spectroscopy data on a series of undoped 
\ce{Hg_{1-x}Cd_{x}Te} ($x$=0.2--0.4) and found two common donor-like recombination centers 
($0.4E_g$ and $0.75E_g$ above the VBM)~\cite{Jones1982}.
Though only scant information was available about these centers, 
Jones \textit{et al.}\ found that their concentrations were proportional to the acceptor concentration, 
thereby tentatively attributing this deep center to Te$_\text{Hg}$ antisite.
This assignment contradicts our calculations indicating that the donor levels of Te$_\text{Hg}$ antisite are slightly below the VBM.
{However, we point out that rather than an isolated Te antisite the defect complex formed in conjunction with a cation vacancy can introduce a deep acceptor level near the midgap.
Nevertheless, the acceptor nature of the complex apparently disagrees with the donor-like defect levels from experiment.
It is possible that extrinsic defects and other defect complexes related to cation vacancies or may contribute to the observed deep donor levels.}


Another possible deep center as pointed out by Littler \textit{et al.}~\cite{Littler1992}
is the Hg interstitial, which shows a direct correlation with the trap level at the midgap.
Experimentally Hg interstitials have been introduced intentionally by thermally decomposition of an anodic oxide layer~\cite{Elkind1992,Littler1992},
and it is plausible that these interstitials could be formed out of equilibrium and 
did not necessarily remain tetrahedrally bonded to the Te atoms as in the low-energy configuration.
Indeed, we have shown that the Hg interstitial introduces a deep level at roughly 0.25$E_g$ above the VBM when tetrahedrally coordinated with 
cations (cf. Figs.~\ref{fig:ctl_int}a and \ref{fig:deeplevels}),
thus providing support for the experimental interpretations.

\begin{figure}
\includegraphics{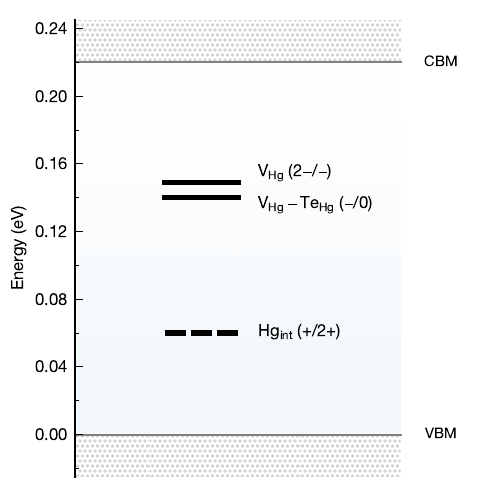}
\caption{\label{fig:deeplevels}Charge transition levels of deep centers in \ce{Hg_{0.75}Cd_{0.25}Te} as identified from the present calculations. The level shown by the dashed line involves a metastable configurations.}
\end{figure}

\section{\label{sec:conclusion}Conclusion}
By means of the DD-CAM dielectric-dependent hybrid functional combined with spin-orbit coupling, we have obtained a comprehensive understanding
of native point defects in LWIR MCT at an unprecedented level of accuracy.
Our results substantiate the role of cation vacancies as the primary deep centers
through the localized $p$-like state from the neighboring Te atoms.
Given its low formation energy and the charge transition level near mid-gap,
we infer that the cation vacancies are likely to play an important role in the SRH recombination
although the exact recombination rate requires further investigation.
{We have also identified Hg interstitial as a possible deep center
through a metastable configuration.
In addition, while Te antisite \textit{per se} is not a deep center,
when it forms a defect complex with a cation vacancy, 
a deep acceptor level can emerge near the midgap.}

The fact that both Hg vacancy and interstitial are deep centers implies that SRH recombinations could
be an issue in MCT for both Hg-rich and Hg-poor conditions. Under Hg-poor growth conditions that is often used for $p$-type doping, the Hg interstitial is less relevant because of its higher formation energy and passivating Hg (and Cd) vacancies might offer a way to limit SRH recombinations. Another possible route would be to perform extrinsic $p$-type doping and bypassing the formation of Hg vacancies, which has been proven to be difficult and demands further research~\cite{Lei2015}. 

\begin{acknowledgments}
This work was supported by the United States Air Force Office of Scientific Research under Award No.\ FA9550-22-1-0355. Computational resources were provided by the supercomputing facilities of UCLouvain (CISM) and Consortium des Equipements de Calcul Intensif en F\'ed\'eration Wallonie-Bruxelles (CECI). Additional resources were provided by the National Energy Research Scientific Computing Center (NERSC),
a DOE Office of Science User Facility supported by the Office of Science of the U.S.\ Department of Energy
under Contract no.\ DE-AC02-05CH11231 using NERSC award BES-ERCAP0023830.

\end{acknowledgments}

\bibliography{main}
\end{document}